# Experience: Quality Benchmarking of Datasets Used in Software Effort Estimation

Michael F. Bosu
*University of Otago and Waikato Institute of Technology, Centre for Information Technology, Wintec, Private Bag 3036, Waikato Mail Centre Hamilton 3240, New Zealand*
email: michael.bosu@wintec.ac.nz

Stephen G. MacDonell
*University of Otago and Auckland University of Technology, School of Engineering, Computer and Mathematical Sciences, Private Bag 92006, Auckland 1142, New Zealand*
email: stephen.macdonell@otago.ac.nz

**Abstract**

*Data is a cornerstone of empirical software engineering (ESE) research and practice. Data underpin numerous process and project management activities, including the estimation of development effort and the prediction of the likely location and severity of defects in code. Serious questions have been raised, however, over the quality of the data used in ESE. Data quality problems caused by noise, outliers, and incompleteness have been noted as being especially prevalent. Other quality issues, although also potentially important, have received less attention. In this study, we assess the quality of 13 datasets that have been used extensively in research on software effort estimation. The quality issues considered in this article draw on a taxonomy that we published previously based on a systematic mapping of data quality issues in ESE. Our contributions are as follows: (1) an evaluation of the "fitness for purpose" of these commonly used datasets and (2) an assessment of the utility of the taxonomy in terms of dataset benchmarking. We also propose a template that could be used to both improve the ESE data collection/submission process and to evaluate other such datasets, contributing to enhanced awareness of data quality issues in the ESE community and, in time, the availability and use of higher-quality datasets.*

## 1. INTRODUCTION

As the name implies, empirical software engineering (ESE) employs observational data in the modelling and understanding of software engineering phenomena. ESE has gained particular prominence in the past decade after Kitchenham Dybå and Jorgensen (2004) espoused the ambitions of evidence-based software engineering, being the incorporation of up-to-date research evidence with practical experience, tempered by human values when making decisions during software development and maintenance. ESE was preceded by "software metrics," a term that referred to activities and data associated with measurement in software engineering. Some of these activities revolved around the production or collection of values to characterize software code properties (the "classic" software metrics) and the development of models to predict various aspects of soft- ware, such as resource requirements, defect rates, broader quality concerns, development effort, and others. The central role of data was evident even then—Hall and Fenton (1997) described soft-ware metrics as including the collection of quantitative measures as a key part of software quality control and assurance activities (and specifically the monitoring and recording of defects during development and testing). This thinking has prevailed. While a range of topics has since been ad- dressed within ESE research, the most substantial bodies of work in this field have proposed or evaluated models constructed primarily for effort/cost estimation or for defect prediction.

The use of metrics in ESE has been asserted as invaluable in facilitating rational decision making during software development and maintenance (Mazinanian et al. 2012; Schalken and van Vliet 2008), with the expectation that this will in turn lead to positive outcomes such as increased development productivity, reduced deployment cycle time, and improved quality of the software product (Daskalantonakis 1992). Although the in-principle benefits of metrics to software engineering is not in doubt, the in-practice benefits have been questioned increasingly in recent years due to growing concerns over the quality of the data being collected and used in the building of models to predict characteristics such as software size and development effort.

The challenges associated with the collection and use of empirical software engineering datasets have thus been documented in several recent publications (Gray et al. 2012; He et al. 2013; Liebchen and Shepperd 2008). Problems such as noise, outliers, and missingness (or incompleteness) have been acknowledged and afforded particular attention by the ESE research community, in terms of both their detection and their resolution (Buglione and Gencel 2008; Khoshgoftaar and Hulse 2005; Liebchen and Shepperd 2008; Liebchen and Shepperd 2005), while other problems, such as poor provenance, inconsistency, and commercial sensitivity, have been largely overlooked. Our previously published taxonomy (Bosu and MacDonell 2013a) identified a number of distinct data quality challenges exhibited in respect to ESE datasets. In this study we apply the taxonomy to some "classic" ESE datasets, found primarily in the PROMISE[1] repository, that have been widely used in studies of software effort estimation[2]. These data sets were selected because they are easily accessible and (so) are frequently used in ESE modeling. Our intent is to benchmark these datasets against the elements of the taxonomy with the goal of evaluating their quality. This will serve to highlight any areas of general concern regarding the collection of ESE data and will also indicate any specific shortcomings in each dataset. We will also gain some insight into the utility of the taxonomy as a benchmarking mechanism. In providing a benchmark of this nature, researchers and practitioners will be able to compare the quality of any new datasets with these classic alternatives. This should lead to more informed decisions as to whether to use a given dataset in ESE modeling.

Although a range of techniques have been proposed to identify or assess the various quality characteristics of ESE datasets, there is no single "front-runner" technique for any of the data quality issues in the taxonomy. As a result, we employ what are considered to be among the best practice technique(s) (described in Section 4) with a view to assessing the quality of these widely used datasets. It is hoped that researchers and practitioners would use appropriate techniques, such as these, in assessing the quality of their own datasets and to in fact develop or utilize new and better methods of data collection; in the meantime, however, the objective of this benchmarking exercise is to illustrate and so promote a holistic assessment of data quality prior to modeling. The contributions of this article are as follows:

- First, we deliver insights into the state of data quality of some of the most widely used datasets in software effort estimation.
- Second, we assess the previously proposed taxonomy in terms of its utility as a mechanism for benchmarking.
- Third, we propose a template that should provide a transparent means of data collection and submission and should support quality assessment of other datasets.

To the best of our knowledge, this is the first study in ESE that has sought to holistically assess the state of quality of a number of commonly used datasets; most prior studies have addressed a limited range of issues or quality concerns associated with one, or perhaps two, datasets. It is also the first study to explicitly advocate the use of a non-proprietary template to guide the collection and submission of datasets to ensure that their quality across multiple relevant dimensions is made clearly "visible."

## 1.1. Motivation

A This work is motivated by previous studies that have addressed the impact of data quality in ESE. In each of the case studies presented in this section, a single data quality issue was addressed. We are of the view that if data quality can be addressed holistically, then ESE practice stands to benefit greatly, given the improvements experienced through the addressing of single data quality issues.

Khoshgoftaar and colleagues applied several noise detection and correction procedures to ESE datasets across a range of studies (Folleco et al. 2008; Hulse et al. 2006; Khoshgoftaar and Hulse 2005; Khoshgoftaar and Rebours 2004), with varying degrees of success. Noise detection techniques, including Bayesian multiple imputation, a clustering-based noise detection approach using the k-means algorithm, an Ensemble-Partition filter, a technique to detect noise "relative to an attribute of interest (AOI)," rule-based noise detection, and Closest List Noise Identification, were applied to various ESE datasets. In all these studies, the authors show that addressing the noise issue in software effort estimation datasets has the potential to improve the performance of their prediction models, leading them to conclude that noise is detrimental to the performance of machine-learning algorithms used in ESE prediction.

Outliers have been a constant source of problems in the analysis of ESE data (Morasca 2009). For instance, Lavazza and Morasca (2012) used a generalized robust regression method to not discard too many data points due to outliers, because as much as 57% of the data points in one of their datasets were determined to be outliers from a Least Squares perspective. The adoption of this approach ensured that they were able to build models that were statistically significant and had superior effort estimation accuracy.

The amount of data available for model building is known to affect the statistical significance of resulting software effort estimation models, with small datasets being particularly challenging. Naive Bayes and Random forest algorithms have been proposed to increase the performance of prediction models based on small datasets and large datasets, respectively (Catal and Diri 2009; Fenton et al. 2008). Another challenge to the amount of data available is missing data values. Zhang et al. (2011) employed two imputation strategies by using the naïve Bayes and Expectation Maximization algorithms to address missingness in software effort estimation datasets. These imputation strategies were applied to the ISBSG and CSBSG datasets, and software effort prediction models were built using the corrected data. The results indicated superior software prediction models based on the corrected data.

---

[1] http://openscience.us/repo/

[2] While "estimation" and "prediction" have slightly differing meanings, in that the latter explicitly refers to the forecasting of a future occurrence, we use the two terms interchangeably here given that many studies actually utilize secondary data sets collected in the past in their analyses (and so, strictly speaking, are analyses of estimation rather than prediction).

It should be evident from the above studies that robustly addressing any aspect of data quality in ESE can lead to improvements in the available dataset and/or the resultant models that are built. Different data quality issues might be associated with different problems. For instance, the presence of noise may mean that a dataset is not fit for purpose, existence of outliers could mean that the results of models might need to be adjusted for skewing, and missing data might lead to the building of models with smaller datasets, which could lessen the power of a model.

The remainder of this article is organized as follows. In Section 2, we present related work, and in Section 3 we describe the datasets selected for assessment. In Section 4, we present the best practice methods used in assessing dataset quality. A discussion of the results of this assessment is presented in Section 5. In Section 6, we propose a template that should aid in data quality assessment and in the collection and submission of datasets in the future. Finally, we present the conclusions of our study in Section 7.

## 2. RELATED WORK

Data are at the core of the practice of ESE, and, as such, its importance to the discipline cannot be overstated. Most researchers use secondary data in ESE modeling (Mair et al. 2005; Shepperd et al. 2014); it is therefore critical that those responsible for collecting data are well trained and aware of the potential problems that could exist in datasets, so that suitable processes are employed to generate, and use, the most reliable data available. At a minimum, the processes used should be documented to inform secondary users of how the data were collected. The challenges faced by those collecting and utilizing empirical software engineering datasets have received increased recognition in recent times (Bosu and MacDonell 2013a; Liebchen and Shepperd 2008; Shepperd et al. 2013), although as a whole the body of literature on ESE data quality remains quite limited (Bosu and MacDonell 2013b). In this section, we review prior assessment studies and we briefly note some of the measures others have taken to improve the quality of ESE datasets and repositories. We first present a representative set of studies that have assessed the state of ESE datasets from one viewpoint or data quality dimension, as this is the predominant approach taken by the ESE community in addressing issues that affect software engineering datasets. In this subsection, we also present example studies that have used metrics from open source projects in building ESE prediction models. This is followed by a review of the few studies that have assessed the state of ESE datasets from multiple viewpoints or considering multiple data quality dimensions.

### 2.1 Single Issue Studies

Noise—erroneous data—has been identified as a problem in several software measurement datasets (Johnson and Disney 1999; Khoshgoftaar and Hulse 2005; Hulse and Khoshgoftaar 2011; Liebchen and Shepperd 2005), and the ESE community has responded with a number of studies seeking to address the incidence and effects of noise. Liebchen et al. (2006) conducted classification experiments to assess the effect of noise on the accuracy of predictions and to evaluate the robustness of techniques for handling noise in ESE datasets. Three noise correction techniques were employed: robust algorithms, filtering, and polishing. Their results demonstrated that polishing is a more effective classification algorithm as compared to robust algorithms and filtering.

Yoon and Bae (2010) proposed a pattern-based outlier detection method that identifies abnormal attributes in software project data and that relies on the existence of normal or typical relationships between attributes, which they termed a data association pattern (DAP). The pattern-based outlier detection method follows a three-step process: First, hierarchical clustering is applied to discretize the numerical attributes of software project data; second, DAPs are mined to identify frequent patterns that meet a certain minimum confidence threshold; and, third, software project data are mapped to the DAPs to identify any abnormal attributes. One of the objectives is to facilitate root cause analysis so as to prevent reoccurrences in the future. The Yoon and Bae (2010) study is significant in the sense that the abnormality of outliers is determined and acted upon relative to other data, in contrast to many studies that classify all outliers as noise and so simply (but perhaps inappropriately) remove them.

Two embedded strategies to address missing data (toleration and imputation) when using naïve Bayes and Expectation Maximization algorithms for software effort prediction were proposed by Zhang et al. (2011). The missing data toleration strategy simply ignores missing values and makes use of existing data values of software projects for prediction. Its strength lies in its low computational complexity requirements. The imputation strategy uses existing values of attributes to estimate missing values. Experimental results drawn from their analyses (of the ISBSG and CSBSG datasets) demonstrated that both strategies outperformed classic imputation techniques.

Inexperienced measurers were identified as contributors of poor data quality in the form of inconsistencies (Cuadrado-Gallego et al. 2010), especially during the data collection stage due to their lack of understanding of software project metrics. It is important for software engineers to be trained in all aspects of data collection so that the quality of the data can be assured.

Redundant and duplicate data in ESE datasets (Bettenburg et al. 2008) might lead to misleading results and can also detrimentally affect the performance of classifiers. Prifti et al. (2011) found that, in their analysis of the Firefox bug repository, there were 748 bugs that had been assigned to multiple groups, after they applied a method that detected duplicates through local references. If effort modeling is based on such data, then clearly there is scope for over-estimation of the actual effort required. Moreover, the building of classification models using data-mining methods will be slowed by the additional processing needed to parse and consider the redundant entries/values.

Models generated from heterogeneous multi-organization datasets have been employed in estimating effort or predicting defects of software projects in a single company in a growing body of research (Bettenburg et al. 2008; Kocaguneli and Menzies 2011; Mendes et al. 2007; Mendes and Lokan 2008; Menzies et al. 2011; Turhan et al. 2009; Zhihao et al. 2005). In spite of the extensive attention given to this issue, results to date have been inconclusive as to whether single organization datasets are superior to those collected from multiple organizations. Kocaguneli et al. (2010) proposed the use of relevancy filtering so that organizations that lack historical data can supplement their software cost estimation with relevant data from other projects or organizations, as this approach was found to be effective as compared to using the data without any relevancy filtering.

The amount of data available for model building contributes to the likely statistical significance of generated models. Small datasets are an acknowledged problem in ESE as they do not lend themselves to the generalization of results. The range of suitable analysis techniques is also con- strained (Bennett et al. 1999; Hall 2007), as some approaches assume the availability of a minimum volume of data. Naturally, this issue is particularly pertinent to organizations that are just beginning a measurement programme or that embark on projects that are substantially different to those undertaken in the past.

Commercial sensitivity is one of several constraints on provenance in ESE. Organizations that hold data that they believe gives them competitive advantage might not be willing to release the data to independent researchers, for fear of proprietary data becoming accessible to competitors. Similarly, they may be reluctant to release data if they believe they could be used to portray them in an unfavorable light. Even when researchers are able to have access to such data, they are often required to sign non-disclosure agreements that prevent them from publishing the data with their results (Liebchen and Shepperd 2005; Mair et al. 2005), thus rendering such studies non-replicable. To resolve the commercial sensitivity problem and promote the sharing of data, Peters et al. (2013) proposed the CLIFF+MORPH algorithm that anonymized data without substantially degrading its use in software defect prediction. This algorithm was applied to good effect on 10 defect datasets from the PROMISE repository.

In the defect prediction study of Turhan et al. (2009), they found it difficult to access failure logs, because several large teams of contractors were working on projects for a single organization— NASA—and each viewed the failure logs as critical to their competitive advantage. The authors note that acquisition of even coarse-grained information was only attained after several years of negotiation. When finally provided, the data were highly sanitized by NASA to the extent that the research team was not able to have information concerning project or module names. Robles (2010) assessed the possibility of replicating experiments reported in papers published in the proceedings of the Mining Software Repositories Workshop/Conference between 2004 and 2009. It was determined that only 6 of 154 experimental papers were replicable, because the data and scripts used in the other 148 original studies were not accessible.

Catal and Diri (2009) performed several experiments to assess researchers' claims that their fault prediction models provided the best performance. When the models were assessed using public datasets, the results were not as strong as had been claimed by their proponents. This may reflect problems with the models themselves (and possible researcher bias), or it may again signal the extent to which models are tied to the underlying data. Whatever the cause, conflicting reports such as this raises trust issues about software engineering experiments and the reliability of the datasets that are used in these experiments.

Empirical software engineering models for effort estimation and defect prediction have been built for open source projects such as the Linux kernel, Mozilla Firefox, Eclipse, and the like. Capiluppi and Izquierdo-Cortázar (2013), in their study of software effort estimation of FLOSS projects using the Linux kernel as a case study, extracted time-aware information from the repository to enable them to identify the occurrence of major development activities. The metrics collected include commits (additions, deletions and modifications), committer, author, major release, timezones (office hours, after office and late night), and code complexity (McCabe's cyclomatic index).

In addition to employing traditional metrics such as function points and Lines of Code, Qi et al. (2017) introduced another group of metrics they termed personal factors, which were determined objectively when they mined the GitHub repository to create effort estimation models for open source projects. The personal metrics are APEX, which refers to the project team experience in a specific kind of application, and LTEX, which is also associated with the programming language and tool experience of the project team. This project was undertaken to address the lack of adequate data for software effort estimation. It is worth noting that personal metrics are not a new idea in software effort estimation datasets, as they have been associated with datasets such as COCOMO, Desharnais, and others.

Metrics such as Commits, PullReqs, PullReqsHandled, ProjectsWatched, IssueComments, IssuesReported, IssuesHandled, Followers, and Mentions have also been tracked in GitHub (Badashian Esteki Gholipour Hindle and Stroulia 2014) to study developer activities. There are therefore di- verse metrics that can be tracked in the GitHub repository for which some can be employed to build effort estimation models, while others have different uses such as commit classifications and developer activity analyses.

Software engineering metrics have also been used to study how software systems change over time. Israeli and Feitelson (2010), for instance, studied the evolution of the Linux kernel. In a 14-year period they considered 810 versions of the system. Some of the metrics used were lines of code, McCabe's cyclomatic complexity, metrics based on Halstead's software science, Oman's maintainability index, Files and directories, and the rate of releasing new versions. Their study found support for Lehman's law in relation to growth and stability of software systems.

Shin, Meneely, Williams, and Osborne (2011) investigated the use of three broad metrics of complexity, code churn, and developer activities as reliable indicators of identifying software system vulnerability. Using the aforementioned metrics, they were able to predict code vulnerabilities in the Mozilla Firefox browser and Red Hat Enterprise Linux.

**2.2 Multiple Issue Studies**

The above studies considered the state of ESE datasets in terms of just one quality dimension; we now consider studies that have assessed the state of ESE datasets from multiple viewpoints.

It has been generally established that the quality of ESE datasets cannot be taken for granted, as data collected even by highly mature organizations can have issues. This is evident in the discovery by Gray et al. (2012) of several data quality problems with the NASA Metrics Data Program (MDP) datasets that are used widely for defect prediction research. The issues evident in these datasets are several and include redundant data, inconsistencies, constant attribute values, missing values, and noise. Shepperd et al. (2013) proceeded further to compare two versions of a NASA dataset (one in the PROMISE repository and the other in the MDP repository) with respect to the data instances and their attributes and discovered that they differed in several respects. They proposed an algorithm that could be used to clean this data of multiple data quality issues.

Rodriguez et al. (2012) used a position paper to classify ESE repositories and the data quality problems that are faced by researchers when using these sources. The repositories were classified into five main groups based on the type of information stored, public or private availability of the dataset, existence of single project or multi-project data, type of content, and the format of data storage. In noting the challenges that these sources posed to (primarily machine learning) researchers the authors referred to difficulties in data extraction, the insufficient provision of information to support replication, and a range of data quality problems, including outliers, missing values, redundant observations, overlapping classes, data shift over time, unbalanced distributions, measurement variability, and model accuracy variability (Rodriguez et al. 2012). The classification of datasets by their distinct properties and the acknowledgement of data quality problems is a positive initiative. The research reported in this article is intended to further enhance data quality in ESE by providing a transparent and consistent means of collection and evaluation that could lead to the use of high(er)-quality data in software engineering experiments.

In a more recent publication, Valverde et al. (2014) proposed a Data Quality model that com- prised data quality dimensions, data quality factors, data quality metrics, and their inter-relationships. Data quality dimensions refer to a broad classification of data quality issues; data quality factors refer to the set of characteristics that makes up a particular dimension; and data quality metrics are the set of measures that are used in assessing the factors in each dimension. The model is intended to support the identification and assessment of quality problems associated with the collection of data from software engineering experiments (Valverde et al. 2014). The authors evaluated the model on two controlled experiments (which compared the effort of developing a web application either by employing a Model-Driven Development approach or a more traditional development approach where code is manually generated). The approach advocated by Valverde et al. (2014) bears some resemblance to the data collection and submission template proposed in Section 6 of this article, as it encourages quality assessment at the data collection stage. Where the two studies depart is that their study considers a subset of the elements of quality that this article considers (specifically, those falling under the Accuracy class of our data quality taxonomy as presented in Section 4). The model of Valverde et al. (2014) also does not directly support independent verification of the data quality issues at stake as it provides only the result of the data quality assessment, whereas the data collection and submission template proposed in this article provides a comprehensive and transparent means of verifying any data collected and all assessments undertaken, with a view to facilitating replication. Such efforts should go some way to addressing quality problems at the data collection stage, which could also be beneficial in terms of early intervention. As noted above, however, empirical software engineering researchers often work with secondary data, and therefore it is similarly important to identify the quality challenges associated with secondary data, a second point of emphasis in this article.

A reasonably recent systematic mapping by Rosli et al. (2013) identified the data quality problem as an issue in ESE and discussed prior assessment techniques as applied to software engineering datasets. Although 10 different data quality problems were identified, nine of them fall into the Accuracy class of the data quality taxonomy (Bosu and MacDonell 2013a). This again signals the sometimes narrow conceptualization of data quality in software engineering, as it is mostly seen from the (albeit important) perspective of accuracy. The present research intentionally adopts a broader conceptualization, and the proposed data collection and submission template should enable users to capture other aspects of data quality in ESE that have to date been largely ignored.

Gencel et al. (2009) attributed the problem of inconsistent results when software effort estimation models are developed using benchmark repositories to two factors:

1. The *lack of common standards and vocabulary*.
2. The *differences in definitions and categories of attributes* of the different repositories.

The authors went on to propose a mechanism for improving the classification of attributes by adapting the parametric estimation method that is used in civil engineering and two software engineering standards (ISO 12182 and ISO 14143-5). The parametric estimating method relies on the use of a classification database of past projects' parameters to estimate new project parameters. The ISO 12182 standard consists of definitions of software application types and the ISO 14143-5 standard is the grouping of software applications into classes based on the functional properties of the software. The authors assert that (more) consistent use of terminology and definitions should lead to better quality ESE data. To date the proposal has been untested. However, in adherence to this suggestion by Gencel et al. (2009), the use of the data collection and submission template proposed in this article should offer a consistent and comprehensive approach for evaluating the quality of data for software engineering experiments.

Cheikhi and Abran (2013) surveyed the PROMISE and ISBSG repositories with the objective of making it easier for researchers to understand the data in them and thus more readily use the data in modeling. The datasets were classified according to the types of studies in which they could be used, such as effort estimation, defect prediction, and others. Properties of the datasets, including the name of the dataset; whether attributes have been described; the source/donor of the dataset; the year the dataset was made available in a repository; and the mode of accessibility of the dataset (such as public or private) were established for each of the data files in the repositories. These important factors form part of the provenance requirement of the proposed data collection and submission template (Section 6).

Denoted "ISBSG" in the preceding text, the International Software Benchmarking Standards Group applies data quality ratings as a mechanism for indicating the quality of the data submitted for inclusion in its repositories:

"This field contains an ISBSG rating code of A, B, C or D applied to the project data by the ISBSG quality reviewers to denote the following:

**A**= The data submitted was assessed as being sound with nothing being identified that might affect its integrity.

**B**= The submission appears fundamentally sound but there are some factors that could affect the integrity of the submitted data.

**C**= Due to significant data not being provided, it was not possible to assess the integrity of the submitted data.

**D**= Due to one factor or a combination of factors, little credibility should be given to the sub- mitted data."

How those heuristics are operationalized in practice, however, is not known. As such, the quality rating has been said to be a proxy for completeness of data (Liebchen and Shepperd 2008), and researchers have tended to discard data with ratings lower than B in their analyses. Using a "blunt" approach such as this may not be optimal, however, in that, depending on the specifics of the research question being addressed, it may be too conservative or too optimistic. In related prior work, we have suggested a more nuanced way to maximize data use from the ISBSG repository (Deng and MacDonell 2008).

The Experience database also uses data quality rating rules, in this case developed by the Finnish Software Measurement Association (FiSMA) to evaluate the data submitted to this particular repository (Forselius 2008). The FiSMA rules are publicly available (Forselius 2008) and anyone interested can apply them to evaluate the quality of data. The rules are designed to ensure that attributes of interest are explicitly described so that all three levels of stakeholders in the data collection process (customer company project management staff, project manager and repository manager) have the same understanding of the data requirements. The FiSMA rules categorize attributes into three classes for which metrics are recorded for each attribute (Forselius 2008). The first class comprises the "basic" attributes of projects such as size, measured in function points, effort, measured in person-hours, and duration, computed from the start date and end date of a project. The second class comprises attributes that are used to determine the context for which projects were developed, such as programming language, platform type, type of projects, and type of business of the customer organization. The third class of attributes are associated with productivity factors of software projects, such as the use of automated tools, customer participation, experience level of developers and project managers, and so on.

There are mandatory attributes (including size of software, effort, start date, end date, and others) for which if any attribute value is missing the data are rejected outright (Forselius 2008). In determining the quality rating of a project, scores are assigned to each attribute, and the scores for all the attributes are aggregated to arrive at a final score for the project. The quality of the content of an attribute value impacts on the score assigned to that attribute. The maximum score possible for a project is 100. In all, seven quality levels are possible upon evaluation of the data. Six of the quality levels are acceptable and mean that records are stored in the Experience database, with the highest data quality level having a score of 90 or above indicated as "AAA" and "D" being the lowest-quality projects stored in the database with scores that lie between 40 and 49. Projects that evaluate to "X" are rejected and not stored in the database. Below are the quality ratings of projects that are assigned based on evaluation of the FiSMA rules by the repository manager:

| AAA | Highest quality | 90+ |
| AA | Excellent | 80–89 |
| A | Very good | 70–79 |
| B | Good | 60–69 |
| C | Satisfactory | 50–59 |
| D | Acceptable | 40–49 |
| X | Rejected | –39 |

FiSMA provides documentation to aid in the determination of scores for the individual attributes—this is said to ensure that the process of evaluating the quality of projects is repeatable and can be carried out independently by all stakeholders. Application of this process of data evaluation is said to have contributed to the increased quality of this repository as compared to the ISBSG repository (Forselius 2008). Project Managers responsible for data collection are also able to use it to self-evaluate the quality of their data prior to submitting it to the Experience database.

In spite of these provisions, some researchers have identified quality issues with this dataset. Outliers and missing and unexplained values have led to the removal of data from this dataset prior to analysis (Maxwell and Forselius 2000; Premraj et al. 2005). Though these problems are acknowledged by Forselius (2008), it is claimed that the Experience dataset is improving in quality upon every new release, due to ongoing enhancement of the rules applied in the collection of the data. Data that do not satisfy the minimum quality requirements are rejected, and so it has been asserted that the Experience database therefore contains high-quality data (Forselius 2008). To indirectly illustrate this focus on quality, the FiSMA rules were applied to the ISBSG dataset in 2008, and it was found that more than 1,000 projects in the ISBSG repository would have been rejected from inclusion if assessed against the FiSMA criteria (Forselius 2008).

The adoption of quality rules in the Experience database is to be commended in terms of con- tributing to improved data collection practices. There are, however, other datasets that have been used in many more ESE studies due to their public availability. While this open availability is positive in terms of facilitating research, we have limited knowledge of how they were collected or of any quality checks that were applied to them—particularly when researchers do not return to the original source of the data. This has motivated us to provide a comprehensive set of data assessment procedures as described in Section 4. Prior to that, we provide a brief overview of the ESE data quality taxonomy (Bosu and MacDonell 2013a), which is the basis of the data benchmarking in this article.

## 2.3 The ESE Data Quality Taxonomy

In this section, we present a brief overview of the ESE Data Quality Taxonomy based on our prior work (Bosu and MacDonell 2013a). The ESE data quality taxonomy was created by surveying a decade of ESE literature on data quality. The study (Bosu and MacDonell 2013a) identified 57 papers that had addressed one or more issues of data quality, and a total of 74 data quality issues were identified by these papers. These issues were grouped into three main classes: accuracy, relevance, and provenance. Sub-issues (or elements) were identified for each of the main classes. We provide a brief definition or explanation of the main issues and sub-issues of the taxonomy as shown in Figure 1.

*2.3.1 Accuracy.* Accuracy refers to the group of data characteristics that, if encountered, renders observed data unfit for modeling. According to the Oxford English Dictionary, accuracy is "the state of being accurate; precision or exactness resulting from care; hence precision ...exactness, correctness." The elements of accuracy are noise, outliers, inconsistency, incompleteness, and redundancy.

*Noise.* Noise is erroneous data or incorrect data—several empirical software engineering studies have identified noise in ESE datasets (Johnson & Disney 1999; Liebchen et al. 2006). Noise is deemed to reduce the accuracy of models; as such, software researchers have proposed noise detection techniques such as Bayesian multiple imputation, rule-based noise detection, and Closest List Noise Identification to address the issue of noise in datasets prior to model development.

*Outliers.* Being data points that lie outside the overall pattern of a distribution (Yoon and Bae 2010), outliers are a common phenomenon in ESE datasets (Johnson and Disney 1999; Liebchen et al. 2006; Yoon and Bae 2010). The presence of outliers might be an indication of an error in the measurement of data or that the data are not fit to be used in the development of a model.

*Incompleteness.* Primarily found in the form of missing values, incompleteness affects several ESE datasets (Liebchen et al. 2006; Liebchen and Shepperd 2005; Chen and Cheng 2006). "Missing" is defined as "not able to be found, because a value is present but not in its expected place, or is not present when it is expected." The definition of incompleteness is, however, broader, as it refers to not complete or finished or imperfect.

It also refers to a part that is not whole or requires some other parts to be complete. Due to the small size of many ESE datasets, the existence of incompleteness in data might render a model statistically insignificant. The ESE research community has proposed several imputation techniques (e.g, Khoshgoftaar et al. 2006; Hulse and Khoshgoftaar 2008, 2014) to deal with the phenomenon of incompleteness/missing data.

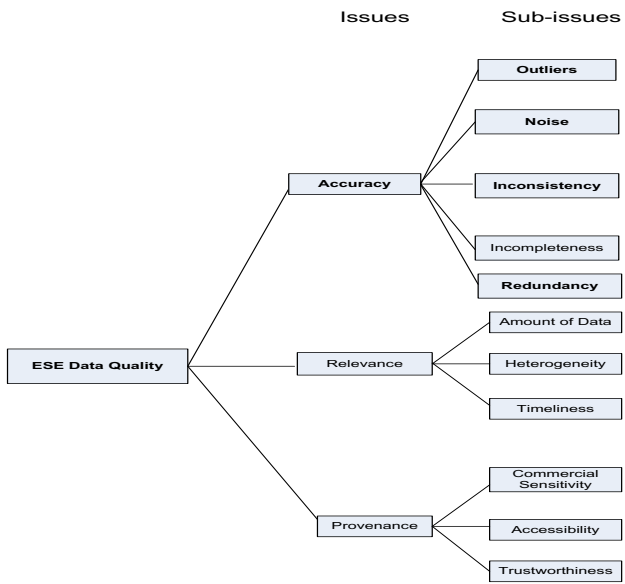

**Figure 1.** Taxonomy of Data Quality in ESE (BosuandMacDonell2013a).

***Inconsistency.*** Inconsistency, according to the Oxford English Dictionary, is defined as "a lack of harmony between parts or elements; instances that are self-contradictory, or lacking in agreement when it is expected." To ensure consistent data in software engineering, it is essential for recorded data to match the variables for which they are recorded. In the study of iterative and incremental software development productivity trends, Tan et al. (2009) discovered inconsistency in effort and size values in that there were mismatches from one report to another.

***Redundancy.*** In software effort estimation and defect prediction datasets, redundancy might exhibit in the form of duplicates or multicollinearity between variables. For example, Prifti et al. (2011) applied a technique that detects duplicates on the Firefox bug repository and discovered as many as 748 bugs that had been described in multiple groups. The use of such a dataset for effort estimation is likely to lead to an overestimation of the required effort—clearly an undesirable outcome.

*2.3.2 Relevance.* The Oxford English Dictionary defines relevance as "the quality or fact of being relevant—bearing upon, connected with, pertinent to, the matter in hand." The use of appropriate data in the development of models—usually classification or prediction is pertinent to the relevance element. Data collected from a different context or system such as real-time embedded system would be unsuitable to be used in estimating development effort for transaction-intensive retail systems. Relevance essentially captures the characteristics of data that are used in modeling. Several ESE studies have considered relevance from the perspective of either single organization datasets or multi-organization datasets. The elements considered under relevance in the taxonomy are heterogeneity, amount of data, and timeliness.

***Heterogeneity.*** In ESE, software effort estimation research has considered heterogeneity particularly in relation to whether the source of the data used in model development is from a single organization or multiple organizations. Researchers have employed heterogeneous datasets from multi-organizations in developing software effort and defects models for single organizations (Mendes et al. 2008; Turhan et al. 2009; Zhihao et al. 2005). Results have proven inconclusive so far as to the superiority of either single organization datasets or multi-organization datasets. The single-company/multi-organization dichotomy may have been oversimplified, as some single organizations are mostly engaged in many diverse projects.

***Amount of Data.*** The statistical significance of models is hugely dependent on the amount of available data used in the development of the models, thus the amount of data available is an important factor of relevance. It is a widely held fact that small datasets is an issue in ESE model development, as they hinder the generalization of results. This also limits the selection of analysis techniques (Bennett et al. 1999; Hall 2007), as some techniques are suited to large amounts of data. Although a dataset might initially consist of large number of records, pre-processing such as the application of stratification and feature set selection approaches could result in data subsets that lack statistical significance power when used in model development. Researchers are therefore required to ensure that pre-processing does not create data subsets that raises questions about results generalizations due to the small nature of datasets and/or the application of inappropriate modeling techniques to the data.

***Timeliness.*** An element of relevance that has received little attention in the ESE research literature is timeliness or currency of data. Mair et al. (2005) conducted a survey in 2005 and found that many ESE studies relied on data that are very old. The analysis of ESE conference and journal publications confirms that these old datasets are still being used in present-day research. To ensure the timeliness of data, it is important for researchers and practitioners to regularly review the characteristics of datasets, taking into consideration the operational context so that the dataset is appropriate for contemporary use. Timeliness is more about the appropriateness of the data use in model development than anything to do with the datasets being inherently "wrong." The question that still remains is "Why are ESE researchers still using old datasets in developing models to be used in effort estimation and defect prediction of contemporary projects?"

*2.3.3 Provenance.* The Oxford English Dictionary defines provenance as "the fact of coming from some particular source or quarter; origin, derivation." The existence of provenance information has been used in the determination of the historical chain of ownership of important objects of value (mostly art work and literature) (Tan 2007). Guaranteeing provenance, while extremely significant for such valuable objects, is also important in relation to results generated by digital systems and other scientific applications. Information regarding provenance establishes something of an audit trail, providing the supporting evidence for scientific results, and, in turn, can directly influence the extent of trustworthiness associated with such results. Because of the reasons enumerated, the value placed on the provenance of digital systems and scientific applications is usually said to be the same as the results they generate (Tan 2007).

Considered broadly, provenance is related to the issue of experimental replication. Replication is or at least should be important (Shull et al. 2008) in empirical software engineering (as it is in other empirical fields) in that it enables the community to build cumulative knowledge concerning which results or observations can be relied upon under different conditions. Shull et al. (2008) advocated the production of good and consistent documentation for all ESE experiments to facilitate replication. This is consistent with previous observation made by Wieczorek (2002), who indicated that a negligible number of empirical software engineering studies were replicated, and, surprisingly, that the use of even the same datasets across multiple studies continued to yield results that were not comparable in most cases, due to differences in the employed experimental designs. She con- tended that the diverse reporting formats of studies in the ESE domain meant that replication and results comparison was a challenge (Wieczorek 2002). The challenge still persists as supported by the Lokan and Mendes (2006) study that replicated cross-company and single-company effort models using the ISBSG database. They were unable to apply the same experimental procedure due to lack of consistent documentation. Replication can be more effective by the use of provenance systems that will provide transparency between the results of an original study and a replicated study.

*Commercial Sensitivity.* Commercial sensitivity is one of the factors that restrains the use of provenance in ESE. This is due to the unwillingness of an organization to disclose and/or release data to researchers outside of their organization when it is believed that the data provide them competitive advantage or might be potentially harmful to the image of the organization. In the rare occasions where data have been released to researchers, they are made to commit to non-disclosure agreements (Liebchen and Shepperd 2005; Mair et al. 2005), which prevents studies based on such data from being replicable. Although non-disclosure agreements protect donor organizations, it limits what can be learned from such data analysis.

*Accessibility.* Researchers having access to data is another issue of provenance in ESE. Turhan et al. (2009) struggled to access the failure logs of NASA due to the fact that several contractors were working on projects for NASA. Each of these independent contractors considered the failure logs as an important element of their competitive advantage. It therefore took several years of negotiations for the researchers to be given access to the failure logs. The released data were highly sanitized to the extent the researchers could not even identify module and project names. Robles (2010) analyzed the experiments reported in published in the proceedings of the Mining Software Repositories Workshop/Conference between 2004 and 2009 with the objective of replicating the studies. To the surprise of the author, a mere 6 of 154 papers were replicable, because there was no access to the data and scripts used in the other 148 papers.

According to Mair et al. (2005), just about 60% of ESE datasets were accessible to the public when the authors investigated the nature and type of datasets that were being used to develop software effort predictions models in the year 2005. Although there have been huge increases in open source development since then, which has made more data available to ESE researchers, it is worth noting that open source development covers several diverse systems with different development practices, which raises questions about its suitability for model development. Another factor is that it could be difficult to map open source model of development to that of industrial software development. To increase the availability of data to researchers, it is essential that public repositories with provenance information such as the ISBSG (www.isbsg.org) and PROMISE (http://openscience.us) should be encouraged. Effective collaboration between academia and industry is another means through which more data can be made accessible to researchers, which ultimately will improve the practice and reliability of ESE models.

*Trustworthiness.* There is a lot of innovation in the field of SE, leading to the creation of new tools, models, techniques, and other related artifacts; however, the field is constrained by lack of rigorous evaluation of these proposals. Glass et al. (2002) concluded that the research approach in SE is narrow and mostly dominated by the "Formulate" approach, with very few studies concentrating on evaluation as a major research activity when the authors analyzed software engineering studies prior to the year 2002. Similar outcomes have been found in other reviews (Clear & MacDonell 2011). It is therefore difficult to have confidence as to the extent to which results reported are applicable beyond the often-limited evaluations performed. This is applicable not only to tools, techniques, and methods; it also affects prediction and classification models as well. Catal and Diri (2009) conducted several experiments to verify researchers' assertions that their fault prediction models provided the highest performance; however, when public datasets were used in assessing some of the models, the results were not as strong as had been claimed by their proponents. This may be due to problems inherent in the models, or it could be an indication as to the extent to which the efficiency of the models is heavily dependent on the underlying data.

Empirical software engineering researchers mostly have limited access to the source of original data, and the most reliable option is to work with secondary data. Researchers therefore have no option but to place their trust in people and systems used in collecting the data and hope that the data are fit for their purposes.

Table 1. Dataset Description

| Dataset | Number of Records | Number of Attributes | Size (unit of measure) | Effort (unit of measure) |
|---|---|---|---|---|
| Albrecht | 24 | 8 | Function Points | Person-Hours |
| China | 499 | 19 | Function Points | Person-Hours |
| Cocomo81 | 63 | 19 | LOC | Person-Months |
| Desharnais | 81 | 12 | Function Points | Person-Hours |
| Finnish | 38 | 9 | Function Points | Person-Hours |
| ISBSG16 | 7,518 | 264 | Multiple | Person-Hours |
| Kemerer | 15 | 8 | KSLOC | Person-Months |
| Kitchenham | 145 | 10 | Function Points | Person-Hours |
| Maxwell | 62 | 27 | Function Points | Person-Hours |
| Miyazaki94 | 48 | 9 | KSLOC | Person-Months |
| NASA93 | 93 | 24 | LOC | Person-Months |
| SDR | 12 | 25 | LOC | Person-Months |
| Telecom | 18 | 4 | Files | Person-Months |

Provenance systems is a means to overcome this challenge, as it will provide researchers and all other data users relevant information, including metadata and the origins of the data, that will increase the trust that is placed in the data. Changes in data, such as masking, anonymization or transformation, and other pre-processing, can subsequently be tracked and verified by both data users and data providers. This is essential in building models with high integrity.

## 3. DATASET DESCRIPTIONS

The 13 datasets used in our quality benchmarking exercise and listed in Table 1 have all been used previously in ESE research (Amasaki 2012; Miyazaki et al. 1994; Prabhakar and Dutta 2013; Shepperd and Schofield 1997). The choice of datasets was informed by a prior study reported in 2005 (Mair et al. 2005) that identified 9 of the datasets as being among those most widely used in software effort estimation, noting that the COCOMO81, Desharnais, Kemerer, and Albrecht datasets were the most widely used of all. The China dataset, although comparatively new (being made available in the PROMISE repository in 2010), has also been included in this assessment, because it consists of 499 records—a large number relative to most other publicly available software engineering datasets. All of the datasets have recently been used together in a number of individual studies by Kocaguneli and colleagues (Kocaguneli et al. 2012; Kocaguneli et al. 2013; Kocaguneli et al. 2015), thus emphasizing their ongoing perceived utility in effort estimation research. An introduction to each of the datasets is provided in this section (in alphabetical order of the commonly used dataset name). Twelve of the datasets have been drawn from the PROMISE repository. (Note that 2 datasets, the Desharnais[3] and Finnish datasets, were available at a previous instance of the PROMISE repository but are no longer accessible.) This study also uses the International Software Benchmarking Standards Group (www.isbsg.org) Release 2016 R1, herein referred to as ISBSG16. All of the datasets contain information reflecting some measure(s) of system size/scope and of development effort, and, as such, these attributes are not emphasized in the description of the datasets. Information concerning the period in which projects were undertaken are stated in the description of datasets where it is known. The number of attributes varies greatly for the datasets—the Telecom dataset consists of only four attributes whilst the IS- BSG16 dataset is composed of 264 attributes. This is noteworthy, as it demonstrates the diversity of datasets and the non-uniformity in the properties collected by different software organizations. In counting the number of attributes, project or record identifiers are also included. While the number of attributes in each dataset varies from just a few variables up to 27 attributes (for Maxwell) and 264 attributes (for ISBSG16), typically only a small number are used in effort modeling.

The **Albrecht** dataset (Albrecht and Gaffney 1983) comprises 24 records collected from projects undertaken in the 1970s at IBM Data Processing Services. The systems themselves were developed using the COBOL, PL/I, and DMS programming languages. System size and complexity were measured using the function point approach proposed by Albrecht (1979).

The **China** dataset comprises 499 projects characterized by 19 attributes. Among these, the function point measures proposed by Albrecht (1979) are again used to quantify system size. It is difficult to provide any further information concerning this dataset—papers that have used this data have provided no background information (Kocaguneli et al. 2013; Prabhakar and Dutta 2013). (An email was also sent to Professor Tim Menzies, who has oversight of the PROMISE repository, and he confirmed that he had no background information on the China dataset and that he received the data in April 2010 without any further details.)

---
[3] At the time of writing the Desharnais dataset can be found at http://promise.site.uottawa.ca/SERepository/datasets/ desharnais.arff

**COCOMO81**, proposed by Barry Boehm (1981), is a software sizing model that has been widely used in the estimation of cost, effort, and schedules for software development projects (Huang et al. 2008). The COCOMO81 calibration dataset used in our assessment is composed of 63 records. It has 19 attributes, including 15 cost drivers that are determined based on the characteristics of the proposed application. The size attribute of the COCOMO81 dataset is measured (or is estimated) in lines of code (LOC).

The **Desharnais** dataset was collected by Jean-Marc Desharnais from 10 organizations in Canada. The projects in this dataset were undertaken between 1983 and 1988. The dataset con- sists of 81 records and 12 attributes, with size measured in function points. In most studies that employ this dataset, 77 of the 81 records are used because of missing data in 4 records (Shepperd and Schofield 1997). In this study, the version that is used in any particular analysis is described as part of the analysis.

The **Finnish** dataset was collected from nine firms in Finland by the TIEKE organization. Initially, 40 records were collected, but missing values in some of the attributes of two projects (Kitchenham and Kansala 1993) meant that their data were removed, leaving 38 records for analysis. This dataset consists of nine attributes, with size measured in function points.

The International Software Benchmarking Standards Group dataset (**ISBSG16**) consists of soft- ware development and enhancement project data collected over several years. This study used Release 2016 R1, which was released in March 2016. The data include project records collected from 32 countries and across more than 12 different industry types (www.isbsg.org). The stated purpose of the ISBSG in compiling the dataset is to aid the software industry in estimating aspects of their projects such as their size, effort, duration, and speed of delivery. The dataset is also said to be useful for benchmarking of projects—so that an organization might compare itself to 'best practice' as represented in the dataset—as well as in the effective planning and management of software projects via software productivity improvements, team size planning, and project risk management. The dataset is available for a fee for commercial organizations. The March 2016 re- lease of the dataset is composed of 7,518 projects with 264 attributes. The size measures used for most of the projects are based on IFPUG function points, but other size measures include NESMA FPs, COSMIC-FP, Mark II FPs, LOC, Dreger, and "Backfired." That said, for reasons discussed later in the paper, researchers often use a subset of the data for modeling, after applying several filters to arrive at the data of interest.

The **Kemerer** dataset (Kemerer 1987) was collected from an American Computer and Consulting firm that developed data-processing software. The data were collected in 1985, and the oldest project at that time was started in 1981, with most of the projects starting in 1983. The projects were said to be medium to large in size based on thousands of source lines of code (measured in KSLOC). The dataset is composed of 15 projects with eight attributes.

The **Kitchenham** dataset (Kitchenham et al. 2002) was collected from American-based multi- national Computer Sciences Corporation (CSC). This dataset contains information related to 145 software development and maintenance projects that CSC undertook for several clients. There are 10 attributes considered, and the size attribute was measured in function points. The attributes also include start date and estimated completion dates, and the projects were undertaken between 1994 and 1999.

The **Maxwell** dataset was collected from a Finnish commercial bank. It is composed of 62 projects represented by 27 attributes (Maxwell 2002). There are 22 categorical attributes that were asserted to have an influence on software productivity. The size attribute was again measured in function points. The start years of projects were between 1985 and 1993.

The **Miyazaki94** dataset was collected by Fujitsu's Large Systems Users Group (Miyazaki et al. 1994). The data were obtained from 48 COBOL systems developed in 20 different organizations and across multiple departments within those organizations. There are nine attributes for each project/system; the size attribute was measured in the number of COBOL source lines of code (in thousands).

The **NASA93** dataset was collected by NASA from five of its development centers (Kocaguneli et al. 2012; Minku and Yao 2013). It comprises 93 projects undertaken between 1971 and 1987. The dataset consists of 24 attributes of which 15 are cost drivers, as the approach is based on that used in COCOMO81. The size attribute was measured in (estimated) lines of code.

The **SDR** dataset was collected from five software organizations in Turkey and is based on the COCOMO II format, having 22 of its 25 attributes as cost drivers (Kocaguneli et al. 2012; Minku and Yao 2013). There are 12 projects in this dataset, and the size attribute was measured in (estimated) lines of code.

The **Telecom** dataset (Shepperd and Schofield 1997) consists of data on 18 software enhancement projects that were undertaken on a U.K. telecommunications product. The version of the dataset used in this study comprises four attributes. Having said that, only the number of files attribute is used in effort estimation, since the other three attributes are not available at the time that estimation would occur.

## 4. DATA QUALITY ASSESSMENT APPROACH

In this section, we provide a description of the methods that we applied to the selected datasets to evaluate them against the taxonomy (Bosu and MacDonell 2013a), which has been briefly described in Section 2.3. The intention is not to develop or promote any particular data quality assessment techniques; rather, the objective is to use known methods to establish the extent to which the data quality challenges identified in the taxonomy may be found in real, widely used ESE datasets. This is important, as we found previously (Bosu and MacDonell 2013b) that data quality assessment is generally not reported in ESE publications. Thus, there is a tendency to simply adopt datasets for analysis without consideration—or perhaps even awareness—of their quality.

Perhaps because of data scarcity, ESE does not use a Kaggle-like approach, wherein datasets are ranked by their users while being made freely available. Much ESE data are proprietary or are closely curated (e.g., by the ISBSG), the Promise repository being the main exception. Under such an approach, the higher-quality datasets would be ranked more highly and so would gain greater prominence, while those with quality problems might lose visibility—and therefore see limited use. While this idea seems appealing in terms of promoting high-quality modelling, it is really moot at present given the limited public access to such datasets.

In the analysis that follows, as many as possible of the available variables and records in each dataset were considered, with the exception of the ISBSG16 dataset, where subsets of attributes and records were used. While the ISBSG16 dataset includes 264 attributes, many records have missing values for a number of these characteristics (due to their not being applicable to a given project, or not being mandatory so not provided by the submitting organization). Therefore, a partial set of the attributes (comprising Functional Size, Summary Work Effort, Development Type, Development Platform, and Language Type) was used in the determination of noise, whilst the (continuous) Functional Size and Summary Work Effort variables were considered in determining outliers. The independent variables selected are known from previous studies to have some degree of influence on effort (Letchmunan et al. 2010; Lokan and Mendes 2009; Seo et al. 2008). Deng and MacDonell (2008) highlighted seven reasons why it might not be possible to use the entire ISBSG dataset for effort estimation, as follows:

- Some variables are not normalized into atomic values.
- Inconsistent recording of variable values.
- There are too many distinct levels for some variables.
- The contexts for some variable values are not discrete.
- Some variables are derived from other variables.
- Some variables are not relevant for effort estimation.
- Some numerical variables have many missing values.

The formalization of the ISBSG release 9 dataset by Deng and MacDonell (2008), with the objective of retaining as many data points and attributes as possible for software project effort prediction, resulted in the identification of 12 usable predictor variables. All of the attributes used here in the assessment of noise and in outlier identification were among those 12 variables. The total number of records retained by the same formalization was 2,862 of the 3,024 records in the ISBSG database. This number in fact represents a substantial proportion of those available, as most studies use fewer than 800 records for modeling. The quality assessment under the three classes of the taxonomy is now presented.

*4.1. Accuracy.* This taxonomy class considers noise, outliers, inconsistency, incompleteness, and redundancy, each of which is now addressed in turn.

**Noise** has been acknowledged as being difficult to determine in respect of ESE datasets (Liebchen et al. 2007), especially when those datasets are secondary sources, meaning the re- searchers may be far removed from their origin. Since it is difficult to be certain about noise in a dataset, and given that researchers may be willing/able to tolerate a certain degree of noise, the assessments undertaken in this study should be interpreted as a guide to the potential state of the datasets rather than definitive statements that a dataset is noisy or otherwise. Even indicative noise assessments such as these are necessary, however, so that researchers and estimators are at least aware of the nature of the datasets they are using and can consider whether preprocessing might be beneficial in improving the quality of the data (and hence any models developed using that data).

Following prior research, we employed two different approaches in determining noise for the 13 datasets selected here. The first approach was to examine whether any formulas used in deriving data were incorrect or violated relational integrity constraints (Shepperd et al. 2013), which are the stated rules/formulas or the expected outcome of a computation. The second technique utilized data classification, where incorrect classification represents a proxy for noisy instances in the data, as implemented by Liebchen et al. (2007). Classification algorithms are able to segment data into the required categories—in this study it is expected that data will be classified as "noisy" or "not noisy." Specifically, for software effort estimation, the classification algorithm identifies a record as noisy where the predicted dependent value of the classifier is different from the actual value. We used a decision tree algorithm (specifically the C4.5 algorithm available as part of the Weka data-mining toolbox) first because it is able to build relationships between data as well as to build models independent of the underlying assumptions of the relationships between the attributes under consideration. Second, decision trees are robust in the presence of missing data, an important consideration given the fact that missingness is a predominant problem in ESE datasets. Third, decision trees are accessible and simple to explain. As such, they have been widely used in general machine learning (González et al. 2008; Moser et al. 2008; Teng 2000) and also in software defect prediction (Folleco et al. 2008; Tang & Khoshgoftaar 2004). Last, to the best of our knowledge, the only two prior studies that attempted to identify or address noisy data in effort estimation datasets (Liebchen et al. 2006; 2007) used the decision tree algorithm. In this article, the effort attribute was discretized, because it was a continuous variable, forming the target class for all of the datasets. Preliminary analyses indicated that most of the datasets could be split into up to four classes; therefore, the discretized effort attribute values were divided into four classes for all 13 datasets. The classifier was then applied to the datasets using fivefold cross-validation. The percentages of the effort class that were incorrectly classified were deemed to be noisy.

Prior to the application of the classifier, a degree of necessary preprocessing was undertaken. Project identifier attributes were removed from the relevant datasets (China, Desharnais, Finnish, Kitchenham, Miyazaki94, NASA93, SDR, and Telecom). As most studies that analyze the Desharnais dataset use the version with 77 projects, the classifier was also applied to this version in our study. One outlier project was removed from the Kitchenham dataset.

Boxplots were generated using the R statistical tool to determine outliers in the datasets under consideration. The plots include the Effort attribute for all 13 datasets as the target outcome variable of interest in software project effort estimation. In general, categorical variables and other attributes that have limited ranges of values were omitted from the plots as follows:

- The *FPAdj* and *AdjFP* of the Albrecht dataset were not included in the boxplot, because there is a transformation relation between them and the *RawFPcounts*.
- In the China dataset, the *Resource* and *Dev_Type* attributes were excluded, because they are categorical variables. *N_effort* was also excluded as it is a transformation of the *Effort* attribute.
- The *LOC* and *Effort* attributes were those plotted for the COCOMO81 dataset, because the other attributes were the cost drivers that are assigned according to a fixed range of values in relation to the application's characteristics.
- For the Desharnais dataset, *TeamExp* and *ManagerExp* were not plotted, because they contain discrete values that range from 1 to 4 and 1 to 7, respectively. *YearEnd* was removed, because it represents project completion date (and so is not known in advance). The *Envergure* and *PointsAjust* attributes have a relation with the *PointsNonAjust* and as such *PointsNonAjust* was plotted, as it has not been subjected to any transformation.
- The *hw* (hardware type), *at* (application type), and *co* (function point contribution of each type) attributes in the Finnish dataset were not plotted, because they are limited-range categorical variables. The *lnsize* and *lneff* are the log transformations of size and effort, respectively, and as such they are not also plotted, since our primary interest lies with original values or attributes (although it is worth noting that the log transformation is a valid preprocessing technique that is often a sensible choice in the case of highly skewed data distributions). The product delivery rate, *prod*, was also not plotted, because it is a derived attribute based on the *effort* and *size* attributes.
- *Size* and *Effort* are the only attributes plotted for the ISBSG16 dataset due to the categorical nature and/or high proportion of missing values for many of the other characteristics. Note that the *Size* and *Effort* records themselves contained several 0s and blanks, and these were removed (leaving a total of 4,805 records) before the boxplots were generated. It is worth noting that the number of records used in the boxplots is higher than that used in the determination of noisy records, because five attributes were considered in finding the noisy instances whilst only two attributes were considered in the generation of the boxplots.
- In the Kemerer dataset, *Language* and *Hardware* are categorical values and as such were not plotted. *AdjFP*, which is a transformation of the *RAWFP*, was also not included on the boxplot.
- In the Kitchenham dataset, the *Start_Date* and *Estimate_CDate* attributes were not plotted, because they represent dates rather than numeric values. The *Client*, *Type*, and *Method* attributes are categorical, and so they were also not plotted.
- In the Maxwell dataset, the *Duration*, *Size*, and *Effort* were the only attributes plotted, because *Syear* represents the start year of projects and the other attributes were categorical variables.
- All the attributes of the Miyazaki94 and Telecom datasets were plotted.
- Only the *LOC* and *Effort* attributes were plotted for the NASA93 and SDR datasets, because the other attributes were categorical and/or had limited ranges of values.

In determining **inconsistency**, we sought original source information about the data and variables so that we could assess the extent to which data might have "moved" from their original state or where questionable and/or repeated values had been included or introduced. Most ESE researchers work with secondary data, and as such we need to be sure we have datasets that are as close to "ground truth" as possible. More than that, if summary statistics were routinely provided with datasets, then this would enable users to check whether the data are likely to be true to the original, as these computations can be quickly performed on other versions of the data (similar to calculating a checksum). More generally, information that accompanies the data in the form of metadata, which explains the relevant details of the attributes and values of a dataset, would seem to be increasingly necessary as it further supports verification of the dataset. Our proposed template is one attempt at promoting the inclusion of such metadata.

**Incompleteness** was relatively easy to determine, as some of the datasets actually state the number of records with missing values. In addition, missing values were represented uniformly as "?" or null values in some of the fields (and such indicators are evident in most ESE datasets). However, when missing values are

Table 2. Results of Noise Classification Assessment

| Dataset | Albrecht | China | Cocomo81 | Desharnais | Finnish | ISBSG16 | Kemerer | Kitchenham | Maxwell | Miyazaki94 | NASA93 | SDR | Telecom |
|---|---|---|---|---|---|---|---|---|---|---|---|---|---|
| Noise (%) | 25.0 | 6.6 | 6.3 | 18.2 | 50.0 | 5.0 | 20.0 | 12.5 | 12.9 | 2.1 | 9.7 | 25.0 | 27.8 |

represented—inappropriately—with zeros (0s), then domain knowledge or metadata are required to interpret such instances correctly.

While the taxonomy considers duplicates and multicollinearity in datasets under the subject of **redundancy**, in this benchmarking exercise only duplicates are sought, because there is no intention at this point to build estimation models with the datasets. We used the advanced filter feature in Microsoft Excel to identify duplicate records. In any case, multicollinearity would be an issue only if certain, related variables were included in a given model. Though multicollinearity is not being given specific attention in this study, we suggest that ESE researchers should routinely ex- amine the correlations among independent variables once they have decided to develop prediction models. This should enable them to avoid introducing the destabilizing effects of multicollinearity in their models.

*4.2. Relevance.* The second of the three classes in the taxonomy considers the amount of data, their heterogeneity, and their timeliness.

A straightforward indication of the amount of data in each dataset was determined by simply counting the number of records; in some cases this information is helpfully stated in the metadata that accompanies datasets. Dataset size is an important consideration in terms of having a sufficient number of records to satisfy the assumptions of the various modeling and analysis methods that are used in effort prediction. In assessing heterogeneity, information on whether the data had been collected from multiple organizations or from a single organization was also sought from dataset metadata. Heterogeneity also relates to other factors, however, such as the different types of application that constitute the projects—data subsets might have distributional characteristics that are distinct from others. It is also worth noting that feature subset selection practices can indicate another form of heterogeneity or the broader aspects of relevance for datasets used in modeling as it selects the variables or features that have the most predictive power instead of using all features for model building. In this article, information on the heterogeneity of datasets was extracted from prior publications that had used these datasets. If a new dataset is donated to a repository and has yet to be reported in a publication, however, then current submission practice and the limited prior reporting of data quality characteristics means that it might be difficult for a researcher or practitioner to know the state of the data with respect to heterogeneity based on its origin. In considering heterogeneity alongside the amount of data, while a single dataset may seem sufficiently large in absolute terms, if it is heterogeneous, then the size of the data subsets becomes another important consideration in terms of their adequacy for analysis.

To benchmark the **timeliness** of the datasets, we determined whether projects were recorded with start and/or completion dates. We used three criteria in determining the era of a dataset from which its general age could be computed:

1. Where start and/or completion dates have been recorded in a dataset the date (year) of the dataset was listed as the range of the earliest project and the latest project recorded. For instance, if the projects in a dataset were noted to have been undertaken between 1998 and 2006, then "1998-2006" was recorded as the year for the dataset.

2. Where start and/or completion dates are not recorded in a dataset, but where there are publications that stated the period in which projects were conducted, the range of the years as indicated in 1 was used to represent the year of the dataset.

3. Where start and/or completion dates are not recorded in a dataset nor stated in a publication, the year of the first publication that referred to the dataset was used as the year of the dataset.

**Provenance**. This third class in the taxonomy considers issues of commercial sensitivity, accessibility, and trustworthiness.

To assess whether datasets faced commercial sensitivity issues, we sought information that might indicate dataset, variable, record, or data item anonymization or transformation. Commercial sensitivity information could also be indicated as part of the metadata embedded in a dataset or provided in a separate document. Since all of the datasets studied here are in public repositories, we deemed all of them to be accessible. In regard to trustworthiness, we sought any documentation that would provide us with detailed information about how and when the datasets were collected, with the intention that the data generation procedure could be checked and/or replicated. Though this detailed information was generally not available,

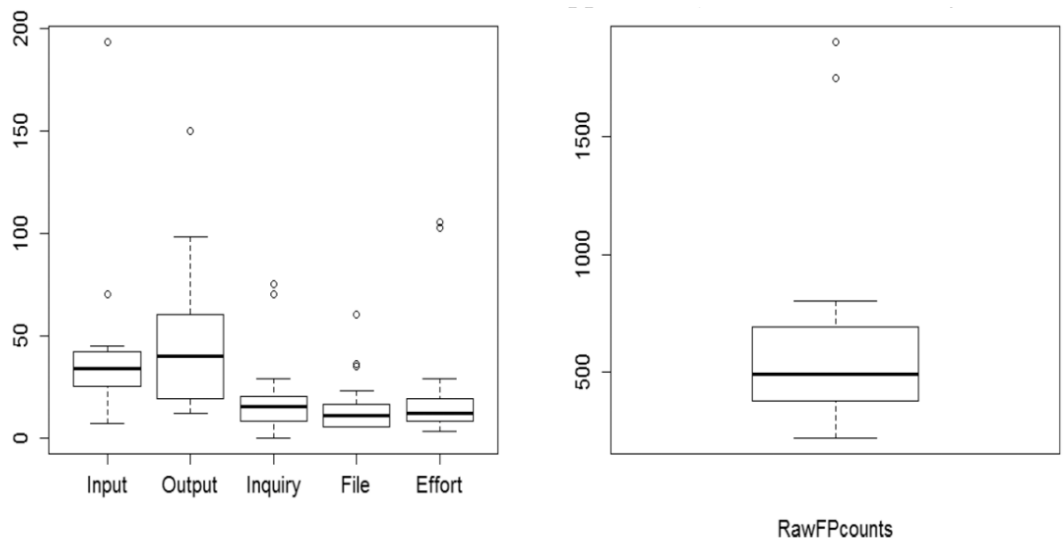

**Figure 2.** Boxplots of Albrecht dataset showing outliers.

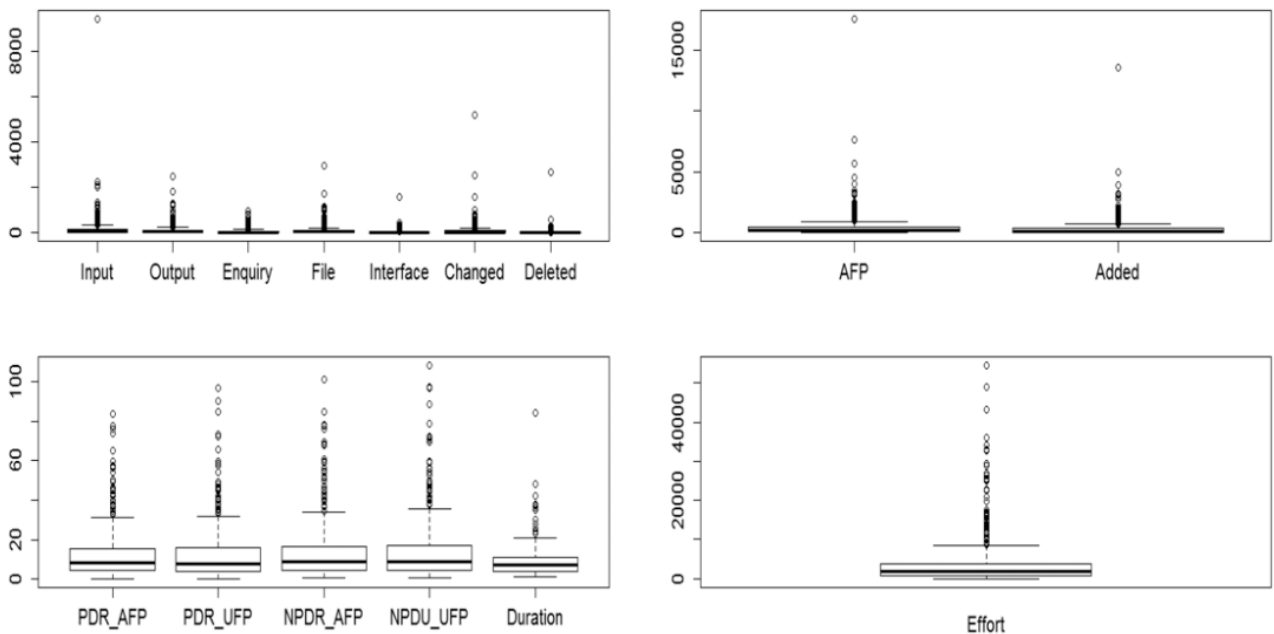

**Figure 3.** Boxplots of China dataset showing outliers.

for some of the datasets there was contact information about the donors of the datasets.

The results of our evaluation of the 13 datasets against the quality criteria in the taxonomy are presented next.

## 5. DATA QUALITY ASSESSMENT RESULTS

**Accuracy Results.** In applying the first criterion used in **noise** determination—that is, revisiting any formulas that were used in generating specific attribute values of a dataset—our analysis suggests that all such formulations were correct, thus implying the absence of noise in the thirteen datasets. However, applying the classification approach with the C4.5 algorithm, where incorrect classification is used as a proxy for noise, yielded incorrect classification rates of between 2% and 50% for the datasets under consideration (as shown in Table 2). Overall, the results in Table 2 indicate an inverse rank relationship between dataset size and noise—the larger datasets tend to be less noisy than their smaller counterparts. Depending on the percentage of the dataset that was incorrectly classified, a researcher might decide to investigate the dataset further, which could result in the dataset either being used or discarded if it would not result in a consistently accurate predictive model.

**Outliers** were evident for at least one variable in all of the datasets, a finding that is consistent with prior literature on this issue that has noted that outliers are a common phenomenon in empirical software engineering datasets (Buglione and Gencel 2008; Liebchen and Shepperd 2005). In particular, there were outliers in the distributions of Effort values for all 13 datasets considered; a subset is shown in Figures 2–4 due to space constraints. The boxplots (Figures I–X) for the other datasets are shown in the electronic appendix. (Note that, for clarity, the boxplots are depicted using different scales.)

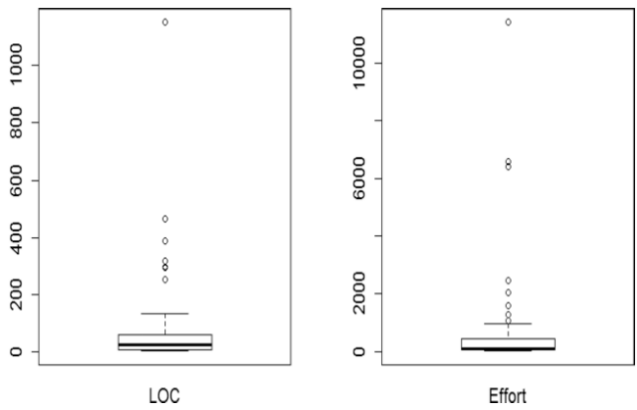

**Figure 4.** Boxplots of COCOMO81 dataset showing outliers.

**Table 3.** Extent of Outliers in the *Effort* Attribute of the Datasets

| Dataset | Albrecht | China | Cocomo81 | Desharnais | Finnish | ISBSG16 | Kemerer | Kitchenham | Maxwell | Miyazaki94 | NASA93 | SDR | Telecom |
|---|---|---|---|---|---|---|---|---|---|---|---|---|---|
| Outliers in the Effort Attribute (%) | 8 | 11 | 14 | 9 | 3 | 10 | 7 | 8 | 8 | 13 | 11 | 17 | 6 |

The percentage of outliers in the untransformed Effort attribute of the respective datasets is shown in Table 3, falling between 3% (Finnish) and 17% (SDR) and bearing no relationship with dataset size. (For this study, data that fall outside the whiskers of the boxplots were deemed to be outliers.) The identification of outliers is important in terms of the reliability of any models generated from a dataset. Researchers and practitioners need to determine the reasons for the incidence of outliers and also employ suitable methods of dealing with the outliers. While not an uncommon practice, it is not appropriate for the outliers to simply be discarded with the only reason being that they are considered noisy without establishing the reasons why those values arose or how their inclusion or exclusion might affect any models generated. Finally, the presence of outliers might also influence the selection of modeling methods (as some, such as robust regression, are more resilient to outlier observations than others).

The Desharnais and ISBSG16 datasets exhibited issues of **inconsistency**. In regard to the Desharnais dataset, questions have been raised over an inconsistency due to the swapping of two variables' labels in some versions of the dataset—PointsNonAjust and PointsAjust were shown above the opposite columns. While their being written in French may have contributed to this occurrence, simple calculations readily made the issue evident and resolvable. Yet researchers have continued to use the wrong data in ongoing analyses of this dataset. In the ISBSG16 dataset, in-consistency was observed in terms of functional size being measured with different units of measurement (NESMA FPs, IFPUG FPs, COSMIC-FFP FPs, Mark II FPs, Backfired, Dreger, Automated, LOC, and Retrofitted). Since data submitters have reported different units for function size measurement, it is the responsibility of those who will use the data for analysis to ensure that they use the right subset of data so as to avoid problems. In addition, implementation date values are not recorded in a uniform format.

In the other datasets, there was no evidence of any inconsistency issues, as shown in Table 4. As a general comment, inconsistency was a challenge to determine, because the information needed was not found in most of the datasets. The provision of provenance information for each dataset would have helped address this situation.

**Table 4.** Results of Inconsistency Assessment

| Dataset | Albrecht | China | Cocomo81 | Desharnais | Finnish | ISBSG16 | Kemerer | Kitchenham | Maxwell | Miyazaki94 | NASA93 | SDR | Telecom |
|---|---|---|---|---|---|---|---|---|---|---|---|---|---|
| Inconsistency | No evidence | No evidence | No evidence | Yes – labels swapped | No evidence | Yes – size measures, date formats | No evidence | No evidence | No evidence | No evidence | No evidence | No evidence | No evidence |

**Table 5.** Results of Incompleteness Assessment

| Dataset | Albrecht | China | Cocomo81 | Desharnais | Finnish | ISBSG16 | Kemerer | Kitchenham | Maxwell | Miyazaki94 | NASA93 | SDR | Telecom |
|---|---|---|---|---|---|---|---|---|---|---|---|---|---|
| Incompleteness | Yes | Yes | No | Yes | No | Yes | No | Yes | No | No | No | No | No |

**Incompleteness** was evident in five of the 13 datasets, which had missing values for some of their attributes, while the remaining eight exhibited no missing data points, as shown in Table 5.

Just over 20% of values in the Inquiry attribute of the Albrecht dataset were missing. Though we found several 0's in some of the fields in the China dataset, we computed missingness in this dataset as reflecting the absence of a value (that is, a blank field), and this resulted in a result of 0.2% missingness for the Effort attribute. This may not be a true reflection of incompleteness in this case, but we are not able to be certain of the meaning of attributes in the China dataset because of a lack of provenance or background information. Five percent (5%) of values in the original 81-record version of the Desharnais dataset were missing, comprising two entries for *TeamExp* and three for *ManagerExp*.

The Kitchenham dataset had missing values in two attributes: About 10% of values in the *Project.Type* attribute were missing, while 2% of the Estimated.completion.date attribute were missing.

**Table 6.** Extent of Missing Values in Selected ISBSG16 Attributes

| Attribute | Extent of Missing Values |
|---|---|
| Summary Work Effort | 0.3% |
| Functional Size | 13.2% |
| Development Type | 0.0% |
| Development Platform | 26.2% |
| Language Type | 18.0% |

**Table 7.** Results of Amount of Data Assessment

| Dataset | Albrecht | China | Cocomo81 | Desharnais | Finnish | ISBSG16 | Kemerer | Kitchenham | Maxwell | Miyazaki94 | NASA93 | SDR | Telecom |
|---|---|---|---|---|---|---|---|---|---|---|---|---|---|
| Number of records | 24 | 499 | 63 | 81 | 38 | 7518 | 15 | 145 | 62 | 48 | 93 | 12 | 18 |

The extent of missingness of selected attributes of the large ISBSG16 dataset is presented separately in Table 6.

An additional attribute (Effort Implement) of the ISBSG dataset that was not of particular im- portance to this study was randomly selected and assessed for missingness, and it was established that close to 78% of its values were missing. This confirms the Deng and MacDonell (2008) study that contended that it would be difficult to conduct software project effort estimation using all the data points in the ISBSG database.

It was also observed that the presentation of missing values was not uniform, in that datasets noted them differently. For instance, in the Kitchenham dataset, missing values were presented as "?," a "0" was used in the Albrecht dataset, "–1" in the Desharnais dataset, and (presumably) a blank in the China dataset. In the ISBSG dataset, missing values were recorded as both blanks and 0s. Understanding missing data points in these datasets would therefore require domain knowledge. There was no redundant data identified in any of the datasets using the method discussed in Section 4. (Note that redundant data points are more prevalent, however, in defect datasets that utilize items such as bug reports.)

**Relevance Results.** The **amount of data** in the datasets varied markedly, between 12 and 7,518 records (shown in Table 7). Four of the datasets comprised fewer than 30 records, which raises a question over whether they could support conclusions with sufficient statistical power if these datasets were to be used in model development (Kitchenham et al. 2002). Moreover, in experiments where splitting of datasets is required (perhaps due to project diversity), this may also result in subsets that are too small to be useful in modeling.

**Heterogeneity** was difficult to determine from the datasets directly, with the ISBSG16 and SDR datasets being exceptions. Although there was no direct evidence provided in or with the other datasets themselves to indicate whether data were sourced from a single company or multiple organizations, we were often able to derive this information from publications that had used these datasets previously. Five of the datasets were collected from multiple organizations (as shown in Table 8), seven were sourced from single organizations, while there was no evidence either from the dataset itself or publications to indicate the heterogeneity status of one dataset—the China dataset. It is worth noting that, although a dataset may be classified as multi-organization, there is the potential for it to contain a significant number of records that belong to a single organization (and an example considered here is the ISBSG dataset that uses a unique (though not visible) ID to identify individual organizations).

The identification of all the single organizations and the total number of records that belong to each would provide an overview of one of the aspects (number of organizations) of data diversity introduced by MacDonell and Shepperd (2007) in their study that compared local and global software estimation models. It would also facilitate further single-company and cross-company analyses.

It should also be noted, however, that the single/multiple organization distinction is a rather simplistic one in terms of being a dominant source of heterogeneity. If we take the Kitchenham dataset used here, for instance, while it was sourced from a single organization (CSC), then the actual projects themselves were undertaken for a wide variety of clients, whose specific contexts, in terms of technologies used, development methods employed, and so on, might mean that other sources of heterogeneity are far more influential in affecting the values for certain data items. Similarly, while the NASA93 projects were indeed all developed for NASA—a single "organization"— five distinct development centers were involved in that work.

**Table 8.** Results of Organizational Heterogeneity Assessment

| Dataset | Albrecht | China | Cocomo81 | Desharnais | Finnish | ISBSG16 | Kemerer | Kitchenham | Maxwell | Miyazaki94 | NASA93 | SDR | Telecom |
|---|---|---|---|---|---|---|---|---|---|---|---|---|---|
| Organizational Heterogeneity | No | No evidence | No | Yes: 10 | Yes: 9 | Yes: Number unknown | No | No | No | Yes: 20 | No | Yes: 5 | No |

**Table 9.** Results of Timeliness Assessment

| Dataset | Dates | Year |
|---|---|---|
| Albrecht | No | 1974-1979 |
| China | No | 2011[P] |
| Cocomo81 | No | 1981[P] |
| Desharnais | Yes | 1982-1988 |
| Finnish | No | 1997[P] |
| ISBSG16 | Yes | 1989-2015 |
| Kemerer | No | 1981-1985 |
| Kitchenham | Yes | 1994-1998 |
| Maxwell | Yes | 1993 |
| Miyazaki94 | No | 1994[P] |
| NASA93 | Yes | 1971-1987 |
| SDR | No | 2000s |
| Telecom | No | 1997[P] |

*[P]: based on when dataset was first published.

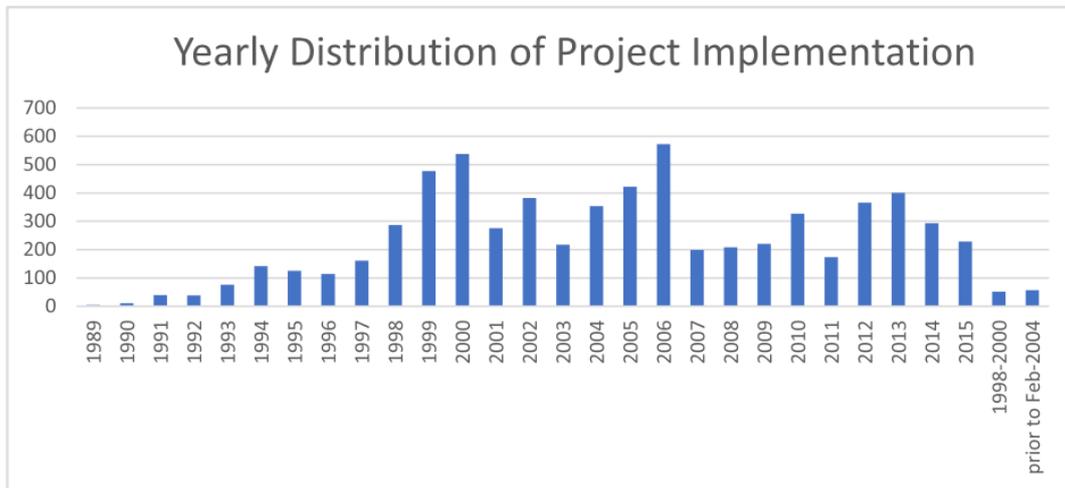

**Figure 5.** Results of Noise Classification Assessment

In considering the criteria for determining data timeliness, only 5 of the 13 datasets contained timing information related to the start date and/or completion date of projects (Desharnais, IS- BSG16, Kitchenham, Maxwell, and NASA93), as shown in Table 9.

Timing information for two further datasets (Albrecht and Kemerer) was derived based on as- sociated publications that provided the start dates and completion dates of the projects (as per the second criterion stated in Section 4). The timing information for a third dataset (SDR), which was also derived based on the second criterion, only specified that the projects were carried out in the 2000s. Information regarding timing for the remaining datasets (China, Cocomo81, Finnish, Miyazaki94, and Telecom) was based on the third criterion described in Section 4—the year the dataset was first used in a publication. The results of the assessment of timeliness are shown in Table 9. Given the large size of the ISBSG16 dataset, Figure 5 further depicts the distribution of projects in that dataset according to their implementation date. Some ISBSG16 projects specified the implementation date as a range: See the "1998–2000" class shown in Figure 5. Date information was also not routinely collected as part of the ISBSG approach until after 2003, hence the "prior to Feb-2004" class. Finally, the implementation date for 78 projects in this dataset are recorded as "Completed." Though on the face of it this might seem intuitive, it is impossible to interpret "Completed" in terms of an implementation date: an illustrative example of inconsistency in the recording of values as we have previously discussed.

**Table 10.** Results of Commercial Sensitivity Assessment

| Dataset | Albrecht | China | Cocomo81 | Desharnais | Finnish | ISBSG16 | Kemerer | Kitchenham | Maxwell | Miyazaki94 | NASA93 | SDR | Telecom |
|---|---|---|---|---|---|---|---|---|---|---|---|---|---|
| Commercial Sensitivity | No evidence | No evidence | No evidence | No evidence | No evidence | Yes | No evidence | No evidence | No evidence | No evidence | No evidence | No evidence | No evidence |

The dynamic nature of software engineering practice would seem to justify that the start and completion dates of projects should be routinely recorded in ESE datasets. This would facilitate analysis related to timeliness, meaning that, for instance, the ESE community would be able to examine longitudinal issues such as productivity variance over time. It would also support the investigation of whether the use of older datasets is relevant to modern day practice.

**Provenance Results. Commercial sensitivity** was generally difficult to determine in any definitive sense in regard to these particular datasets, as no information had been provided regarding portions of the data being hidden or anonymized. The ISBSG16 dataset was the only one that explicitly reflected the issue of commercial sensitivity in the accompanying field description document (and as implemented through the randomizing of project IDs and the removal of any relationship between projects and organizations). That said, for eight of the datasets (excepting Albrecht, COCOMO81, Kitchenham, NASA93, and SDR) there is no information relating to the names of the organizations that collected and/or donated the data. The results of the commercial sensitivity benchmarking evaluation are shown in Table 10.

Since the 13 datasets under consideration here are in a public repository, we deemed all of them to be **accessible**. In contrast, datasets such as the Experience Database and Tukutuku (Mendes et al. 2008), which have been used in some ESE studies, are not in the public domain and so would not be considered as accessible. There are further unknown and unavailable datasets (Abrahamsson et al. 2011; Lee et al. 2014) that have been used in previous ESE studies but have not been considered here.

Table 11. Results of Provenance-Trustworthiness Assessment

| Datasets | Albrecht | China | Cocomo81 | Desharnais | Finnish | ISBSG16 | Kemerer | Kitchenham | Maxwell | Miyazaki94 | NASA93 | SDR | Telecom |
|---|---|---|---|---|---|---|---|---|---|---|---|---|---|
| Provenance-Trustworthiness | Yes | No | No | Yes | No | Yes | No | No | No | No | Yes | Yes | No |

Only five of the datasets (Albrecht, Desharnais, ISBSG16, NASA93, and SDR) provided any sort of provenance information (as shown in Table 11), although we consider this to be a minimal set, as it might not lead to the successful reproduction of such data (as it was mostly oriented to recording the contact/donor information for datasets).

The ISBSG uses a questionnaire to collect data (www.isbsg.org), which might be useful in enabling a more reliable and repeatable process of data collection. The questionnaire provides sections for collecting detailed information from data submitters—although the ISBSG keeps this in- formation confidential. Information concerning the project process, comprising all the activities that took place during a project and the technology used for a particular project, are also recorded. The work effort expended by the people involved in a project is also recorded although no personal information is collected. Detailed information about the software product or application created and the functional size of the software are also recorded. Data concerning the entire project are recorded when a project is completed. Organizations that use the ISBSG questionnaire for collecting data about their projects can develop procedures for auditing the data collection process, which could lead in principle to an increase in the **trustworthiness** of the data collected.

## 6. TOWARD MORE EFFECTIVE DATASET COLLECTION AND SUBMISSION

In conducting the above analysis, it became apparent that it is not uncommon to find inconsistencies in the recording and reporting of ESE datasets—such as different studies reporting different numbers of attributes for the same dataset, differences in record numbers, and different names for the same dataset and/or the variables in it. The routine provision of provenance information, coupled with the use of the template proposed in this section, could address some of these problems. We provide a number of examples here—note that our intent is not to claim one source to necessarily be "more correct" than another but to simply highlight the prevalence of inconsistent reporting.

Azzeh et al. (2010) reported the number of attributes in the Albrecht dataset to be 7, although this is contrary to the dataset in the PROMISE repository, which contains eight attributes and as per the original dataset shown in the first publication that used the Albrecht dataset. Two studies, Huang and Chiu (2009) and Reddy and Raju (2009), reported the COCOMO dataset to consist of 17 effort drivers, which is contrary to both what was reported by Nguyen et al. (2008) and the dataset that is in the PROMISE Repository (which consists of 17 attributes in total of which 15 are cost or effort drivers). Tosun et al. (2009) reported the Desharnais dataset to consist of 10 features, although the dataset used in this study is composed of 11 attributes, in line with what was reported in Desharnais' thesis (Desharnais 1988) and also by Li et al. (2009). Banker et al. (1994) reported the number of records in the Kemerer dataset to be 17, which is contrary to the 15 we have sourced from the repository and as also noted by other researchers (Shepperd and Schofield 1997). Hsu and Huang (2007) reported the number of features of the Kemerer dataset to be 6, though 7 was originally reported. Though the Finnish dataset used in this study is composed of 38 records, which is the same as has been previously reported (Shepperd and Schofield 1997), it was reported by Kitchenham and Kansala (1993) as consisting of 40 projects. Though several publications refer to the Kitchenham dataset as CSC (Amasaki et al. 2011; Amasaki 2012; Keung and Kitchenham 2008), the PROMISE repository refers to it as the Kitchenham dataset (as also used in this study). The Finnish dataset has been variously known as the Laturi, STTF, and initial Experience dataset (MacDonell and Shepperd 2007). Clearly it becomes challenging to identify a dataset if it is referred to using different names and the appropriate provenance information has not been kept. Though the Desharnais dataset was collected from 10 different organizations, some studies refer to it as coming from a Canadian software house, giving the impression that it is a single-company dataset (Tosun et al. 2009), which could lead to it being used wrongly in comparisons of single-company and multi-company analyses.

Advances in science typically rely in part on replication—the construction of a compelling body of consistent evidence through a series of independent tests. Such tests are only possible, however, when sufficient detail is provided to enable faithful replications to be conducted. In this respect, the provision of ESE datasets for research needs to be augmented by provenance information, so that researchers can readily verify the data they intend to use in modeling, or they can make an informed decision not to use certain data in modeling. Reflecting on the reporting inconsistencies presented above, and the issues that we encountered in our benchmarking exercise, we propose a template that could be used to accompany the collection and submission of datasets to public repositories with the objective of ensuring that such datasets are collected, submitted, and used in an informed and consistent manner by ESE researchers and practitioners. The template in Table 12 is intended to address this need by providing a means through which the nature and origin of an ESE dataset will be more transparent to its users. Adoption of this template (or something similar) should also provide support for the explicit identification (and perhaps the resolution) of data quality issues, as far more information about datasets will be provided than has typically been the case to date.

Finally, it should also enable researchers and data collectors to adapt and improve the methods they use in collecting data, as they will be more aware of the challenges that can arise in relation to data quality. The overall objective of the template is to provide a uniform record to support data collection, submission, and use.

In a related study, Mair et al. (2005) collected and reported information relating to ESE datasets from research papers published until 2004. The information collected included dataset name, version, public availability, contact person, start and completion dates, nationality, number of organizations, application domain (business sector), number of projects, project type, number of features, and missing values. A further study (involving the second author of this work) (MacDonell and Shepperd 2007) also classified datasets used in effort modeling according to the following criteria: data quality, including collection and verification; completeness and whether the submission of data had been incentivized in any way; and data diversity, including countries of origin, organizations of origin, and the targeted application domains.

The template proposed above contains some but not all of the properties collected by these studies, in line with their different objectives. The intent of the Mair et al. (2005) study was to assess and characterize the types of datasets that were used in software project effort estimation. MacDonell and Shepperd (2007) evaluated a group of datasets in the context of their study of single- versus multi-organization predictions of development effort. In contrast, the goal of our template is to ensure that detailed information is provided with all datasets so that users can more readily assess the quality of the data as well as to increase the trust that is associated with various datasets. It is also intended as a means of providing uniform guidance in terms of which data should be collected and submitted to repositories where possible.

Contributors of datasets who provide information concerning noise, outliers, inconsistency, incompleteness, redundancy, and the total number of records as stipulated by the template are also providing users of datasets with an opportunity to verify the correctness of those datasets. Where a discrepancy exists in dataset versions, users will be able to contact the right person to remedy this, using the information that fully addresses the dataset's provenance. This should help to support more extensive replication of ESE data analyses.

The provision of heterogeneity information should mean that the number and (possibly anonymized) identity of the organizations that contributed to a dataset are known. It should also provide information about factors that might be used to group projects, such as the type of ap- plication developed or the industry sector(s) that is meant to use the application. Information concerning relevant application and industries types is useful for organizations in benchmarking their datasets for similar applications and industries.

Provision of timing information would ensure that start dates and completion dates are recorded for projects or within-project activities. This would enable the derivation of the duration of projects and would also offer the opportunity to model effort prediction over time.

Mair et al. (2005) noted that much of the data used in empirical software engineering studies were at that time not publicly available. If commercial sensitivity can be more effectively managed, then this would offer the ESE community the opportunity to address issues that will make it more attractive (or at least more acceptable) for more organizations to make their data available for research. While the availability of repositories such as the ISBSG, PROMISE, and those comprising numerous open source projects might have been expected to lead to greater openness and more publicly available datasets for use in ESE studies, our own earlier study (Bosu and MacDonell 2013b), which reviewed empirical software engineering papers published between January 2007 and September 2012, found that still a third of the datasets used were not in the public domain.

Any problems encountered during data collection, if known and reported, should inform more justified use of the resulting data, as well as the potential development of better data collection methods. Inclusion of provenance information would provide the detail necessary to enable the replication of a data collection process. The recommendation of collecting provenance information as part of the proposed template is not intended to ignore the privacy and commercial sensitivity concerns of data submitting organizations. It is rather to ensure that repository managers can turn to owners of data when they discover challenges with data to facilitate easy and timely resolution of data quality problems. Research considering the possible application of techniques such as masking, transformation, and normalization of data, while retaining the integrity of the data values and the relationships between records and attributes, could underpin new techniques that organizations could use to "confidentialize" their data, lending them reassurance around its submission.

Table 12. Template for Dataset Collection/Submission

| Data Quality Challenge | Parameters/Information |
|---|---|
| Noise | 1. Formulas used in generating derived attributes<br>2. Number and proportion of records correctly/incorrectly classified<br>3. Method/tool used to assess noise |
| Outliers | 1. Attributes with outliers<br>2. Identifiers of records with outliers<br>3. Number and proportion of outliers<br>4. Method/tool used to identify outliers |
| Inconsistency | 1. Total number of attributes/variables<br>2. Detailed explanation of attributes and their measurement<br>3. Range of values for each attribute<br>4. Summary statistics for each attribute |
| Incompleteness | 1. Attributes with incomplete data<br>2. Identifiers of records with missing values<br>3. Total missing data points for each attribute<br>4. Number and proportion of incompleteness<br>5. Reasons for incompleteness |
| Redundancy | If any redundant data exists in a dataset—what are the reasons? |
| Amount of Data | 1. Total number of records |
| Heterogeneity | 1. Number (and name where possible, or ID) of organizations/groups that contributed data<br>2. If heterogeneous—number of projects from each organization/group<br>3. If heterogeneous—identifiers of projects in each group<br>4. Type of industries that will use the software for each project |
| Timeliness | 1. Start date of project<br>2. Completion date of project<br>3. Distribution of project effort over time |
| Commercial Sensitivity | 1. Attributes that have been pre-processed or removed because of need for anonymity<br>2. Reasons for commercial sensitivity |
| Accessibility | List of problems encountered during data collection |
| Provenance | 1. Organization(s) from which data were collected.<br>2. Organization that collected the data<br>3. Contact details of person responsible for data collection<br>4. Purpose of data collection<br>5. Software development methodology used for each project<br>6. Data collection method<br>7. Data pre-processing techniques<br>8. Contributor[s] or Donors of dataset<br>9. Date of collection<br>10. Name of dataset<br>11. Version |

Where there is information that is not clear about a given dataset, the relevant contact information of the dataset collector would be available. More generally, use of the template should help to ensure that organizations that are submitting high-quality data are known—and their data collection methods and procedures could then be adopted by others to improve the general state of ESE data.

Though the adoption of the template might increase the workload of software engineering professionals involved in data collection, it is contended here that most of the required information is already available—it is simply not being recorded and/or submitted at present. To continue to improve empirical software engineering as an evidenced-based discipline, more effort along the lines just described should be exerted in supporting the transparent collection and sharing of high- quality data.

We acknowledge that a ranking system, in which a weighted value is assigned to each data quality issue, could be an appropriate means to determine and represent the quality of ESE datasets. The state of data quality practice in ESE has not matured to this extent; however, we believe that this study and the proposed data quality template are a first step in providing some measure of objectivity in the selection of datasets for ESE modeling.

## 7. CONCLUSION

In this study, we have applied a range of data quality assessment techniques to 13 widely used ESE datasets with the objective of benchmarking them against the taxonomy proposed by Bosu and MacDonell (2013a). The issues were addressed one by one in our analysis, and the overall results of this exercise are summarized in Appendix A. It is evident that these datasets do not contain sufficient information to enable researchers to identify any inconsistencies, commercial sensitivities, and their provenance. Timing information was also not provided in most cases with these datasets. Considering the fact that software engineering is a dynamic discipline, it would seem to be imperative that timing information, such as the beginning and completion dates of projects, is provided with ESE datasets. This would enable researchers and practitioners to build models over time, thus supporting assessments of the impact of the adoption of new development techniques, for instance. It was also challenging to determine whether datasets were collected from a single organization or multiple organizations in several cases. Since there is still a degree of contention about the superiority of models generated with either dataset type, it would be appropriate if this information was included with datasets that are provided for modeling.

Techniques have been developed by the empirical software engineering research community to address challenges such as outliers, incompleteness, and, to some extent, noise in datasets. Aspects of data quality that have received far less attention from the community are commercial sensitivity, inconsistency, and provenance. Use of the template proposed in Section 5 would address this lack of attention, providing a transparent means of collecting, submitting, and assessing the quality of a dataset.

## ACKNOWLEDGMENT

The work of M.F. Bosu was supported by a University of Otago Doctoral Scholarship. M.F. Bosu also thanks the Graduate Research Committee of the University of Otago for supporting this research with a Postgraduate Publishing Bursary (doctoral).

## APPENDEX A

Table 13. Summarized Results of Dataset Quality Assessment

| Datasets | Noise | Outliers | Inconsistency | Incompleteness | Redundancy | Amount of data | Heterogeneity | Timeliness Dates | Timeliness Year | Sensitivity Commercial | Accessibility | Provenance/ Trustworthiness |
|---|---|---|---|---|---|---|---|---|---|---|---|---|
| Albrecht | 25.0% | Yes | No evidence | Yes | No | 24 | No | No | 1974-1979 | No evidence | Yes | Yes |
| China | 6.6% | Yes | No evidence | Yes | No | 499 | No evidence | No | 2011[P] | No evidence | Yes | No |
| Cocomo81 | 6.3% | Yes | No evidence | No | No | 63 | No | No | 1981[P] | No evidence | Yes | No |
| Desharnais | 18.2% | Yes | Yes | Yes | No | 81 | Yes | Yes | 1982-1988 | No evidence | Yes | Yes |
| Finnish | 50.0% | Yes | No evidence | No | No | 38 | Yes | No | 1997[P] | No evidence | Yes | No |
| ISBSG16 | 5.0% | Yes | Yes | Yes | No | 7518 | Yes | Yes | 1989-2015 | Yes | Yes | Yes |
| Kemerer | 20.0% | Yes | No evidence | No | No | 15 | No | No | 1981-1985 | No evidence | Yes | No |
| Kitchenham | 12.5% | Yes | No evidence | Yes | No | 145 | No | Yes | 1994-1998 | No evidence | Yes | No |
| Maxwell | 12.9% | Yes | No evidence | No | No | 62 | No | Yes | 1993 | No evidence | Yes | No |
| Miyazaki94 | 2.1% | Yes | No evidence | No | No | 48 | Yes | No | 1994[P] | No evidence | Yes | No |
| NASA93 | 9.7% | Yes | No evidence | No | No | 93 | No | Yes | 1971-1987 | No evidence | Yes | Yes |
| SDR | 25.0% | Yes | No evidence | No | No | 12 | Yes | No | 2000s | No evidence | Yes | Yes |
| Telecom | 27.8% | Yes | No evidence | No | No | 18 | No | No | 1997[P] | No evidence | Yes | No |

*P-based on when dataset was first published.